\documentclass[a4paper,10pt,amsmath,amssymb,notitlepage,twoside,nofootinbib,floats,floatfix,eqsecnum,aps,prd]{revtex4}
\usepackage[latin1]{inputenc}
\usepackage[T1]{fontenc}
\usepackage[usenames]{color}
\usepackage[final]{showkeys} 
\usepackage[pdftex]{hyperref}

\renewcommand{\ge}{\geqslant}
\renewcommand{\le}{\leqslant}

\def\l{\lambda}
\def\om{\omega}
\def\p{\partial}
\def\L{\Lambda}

\makeatletter\g@addto@macro\bfseries{\boldmath}\makeatother%

\def\0{\nonumber}
\def\lb{\label}
\def\ie{i.e.}
\def\eg{e.g.}
\def\be#1\ee{\begin{align}#1\end{align}}
\def\bsube{\begin{subequations}}
\def\esube{\end{subequations}}

\begin{document}

\title{Asymptotically flat black holes sourced by a massless scalar
field}

\author{Mariano Cadoni}\email{mariano.cadoni@ca.infn.it}
\author{Edgardo Franzin}\email{edgardo.franzin@ca.infn.it}
\affiliation{Dipartimento di Fisica, Universit\`a di Cagliari\\
             \& INFN, Sezione di Cagliari\\
             Cittadella Universitaria, 09042 Monserrato, Italy.}
\date{\today}

\begin{abstract}
We derive exact, asymptotically flat black hole solutions of 
Einstein-scalar gravity sourced by a non 
trivial scalar field with~$1/r$ asymptotic behaviour.
They are determined using an ansatz for the scalar field profile and
working out, together with the metric functions, the corresponding
form of
the scalar self-interaction potential. 
Near to the singularity  the  
black hole   behaves as the Janis-Newmann-Winicour-Wyman  solution. 
We also work out a consistent  thermodynamical description  of our 
black hole solutions. For large mass our hairy black  holes  have  
the same thermodynamical behaviour of the Schwarzschild black hole, 
whereas for small masses they differ substantially from the latter.

\end{abstract}

\maketitle

\section{Introduction}
In the past, the issue of the uniqueness of the Schwarzschild black 
hole has motivated the formulation of  
no-hair theorems~\cite{Israel:1967wq,Bekenstein:1995un}  forbidding 
the existence of black hole solutions endowed with a non trivial
scalar 
field. Later, it was discovered  that some low-energy string models
allow 
for black hole solutions with scalar
hair~\cite{Gibbons:1987ps,Garfinkle:1990qj,Cadoni:1993yt,Monni:1995vu,Duff:1999gh}.
Nevertheless, the existence of these solutions remains  limited to
gravity 
theories with non-minimal couplings between  the scalar field and the 
electromagnetic field.

Recently, the application of the AdS/CFT correspondence  to condensed 
matter systems  generated a flurry of activity on  the search of new
black 
hole and black brane solutions with AdS or domain
wall~\cite{Cadoni:2011yj,Cadoni:2012uf,Cadoni:2012ea} asymptotics 
endowed with scalar
hair~\cite{Gao:2004tu,Hartnoll:2008vx,Horowitz:2008bn,Hartnoll:2009sz,%
Mignemi:2009ui,Cadoni:2009xm,Horowitz:2010gk,Cadoni:2011kv,Cadoni:2011nq}.
The main 
reason behind this interest is the holographic interpretation  of the 
scalar field. In the dual  QFT  the scalar field  plays the role of 
an order parameter whose non trivial profile  generates symmetry 
breaking and/or phase transitions.
Shifting  to asymptotically AdS solutions allows  to circumvent the
standard
no-hair theorems, relating  the existence of hairy black holes 
to the violation of the positive energy theorem
(PET)~\cite{Torii:2001pg,Hertog:2006rr}.
In the AdS spacetime, differently from the flat case, a scalar field 
may have negative squared-mass, without
destabilizing the vacuum~\cite{Breitenlohner:1982bm}.

Consequently, several numerical and analytical, black hole and black
brane, 
solutions with scalar hair   have been
derived and used in the literature  for holographic
applications~\cite{Hartnoll:2008vx,Horowitz:2008bn,Hartnoll:2009sz,Horowitz:2010gk,Cadoni:2009xm,Cadoni:2011nq,Cadoni:2011kv,Gao:2004tu,Mignemi:2009ui,Cai:1996eg,Charmousis:2009xr,Gouteraux:2011ce,Cadoni:2013hna}.

In spite of this advances on the side of  black hole solutions with
AdS 
asymptotics, very few progress has been achieved in the 
search for asymptotically flat (AF) black holes with scalar hair
(see, however, Refs. 
\cite{Herdeiro:2014goa,Herdeiro:2014jaa,Charmousis:2014zaa,Anabalon:2012ta,Anabalon:2012ih,Anabalon:2013baa,Anabalon:2013qua}).
On the other hand, it is becoming increasingly  evident that scalar 
fields  play a key  role for understanding the  physics of 
fundamental interactions.
The recent  discovery of the Higgs particle at LHC has confirmed 
experimentally the existence of a fundamental scalar particle 
responsible for the breaking of the electroweak
symmetry~\cite{ATLAS:2012ae,Chatrchyan:2012tx}.
Observations of the Planck 2013 satellite give a striking 
confirmation of  the existence of cosmological inflation generated by 
a scalar field coupled to gravity~\cite{Planck:2013jfk}. Finally,
the present accelerated expansion 
of the universe could be also explained by Einstein gravity coupled 
to  a scalar field.

An other important point to be mentioned  here, is the universal 
scaling behaviour  observed, numerically, years ago  by Choptuik in
black hole formation due to the collapse of a scalar field
~\cite{Choptuik:1992jv}.

It is therefore of  considerable interest to ask about the 
relevance of non trivial scalar field configurations  for 
asymptotically flat black holes.
In this paper we  consider the simplest case of asymptotically flat
black holes 
sourced by a massless scalar field $\phi$. Since at high energies the
mass 
term can be neglected with respect to the kinetic term, we expect  
this situation to describe the short scale behaviour of any 
Einstein-scalar theory of gravity.

Static, spherically symmetric, AF, gravitational solutions sourced by
a non 
trivial  scalar field with an 
identically vanishing potential~$V(\phi)$  are known since a long
time~\cite{Janis:1968zz,Wyman:1981bd,Virbhadra:1997ie}.
They are called the  Janis-Newmann-Winicour or  Wyman (JNWW) 
solutions.
Consistently with the PET they do not describe black holes  but naked 
singularities. 
To find AF hairy black hole solutions one has to relax the condition 
$V=0$. We will therefore consider a potential which is zero only 
asymptotically. Moreover, in order to keep the scalar massless we 
will impose the $r\to\infty$ fall-off behaviour for the 
scalar, $\phi\sim 1/r$. We will show that this boundary condition 
constrains in a highly non trivial way the asymptotic~$\phi=0$
behaviour of~$V(\phi)$. 

Exact AF hairy black hole solutions are then derived and written in
closed  form by 
using the solution-generating method proposed in
Ref.~\cite{Cadoni:2011nq}, \ie\ we derive the 
potential $V(\phi)$ by assuming a profile  
for the scalar field
consistent with the $r=\infty$ boundary condition. 
The potential turns out to be  
 unbounded from below.  This rather unpleasant feature seems 
however unavoidable in view of the PET\@.

Our black hole solution reduces to the JNWW one near to the
singularity.  
We  then investigate in detail its thermodynamics and we show that in
the large mass limit it behaves as the Schwarzschild black hole.

The structure of this paper is  as follows.
In Sect.~\ref{sect:s1} we present the general Einstein-scalar theory 
of gravity we consider and  briefly review the solution-generating 
method of Ref.~\cite{Cadoni:2011nq}. In Sect.~\ref{sect:sss} we
rederive the JNWW 
solutions and discuss their main features. Boundary conditions on the 
scalar field and the corresponding asymptotic behavior  for $V(\phi)$ 
are discussed in Sect.~\ref{sect:ab}.
Our hairy black hole solutions are derived and discussed in
Sect.~\ref{sect:bhs}.
The thermodynamical behaviour of our solutions is discussed in
Sect.~\ref{sect:thermo}.
Finally, in Sect.~\ref{sect:con} we present our conclusions.

\section{Einstein-scalar gravity  sourced by a  scalar
field\lb{sect:s1}}

We consider Einstein gravity in four spacetime dimensions minimally
coupled with a scalar field~$\phi$:
\be%
A=\int{}d^4x\sqrt{-g}\left(\mathcal{R} -2 (\p\phi)^{2}-V
(\phi)\right)\lb{action},
\ee%
where $\mathcal{R}$ is the scalar curvature of the spacetime and we
use natural units with $G=1/16\pi$.

The field equations are
\bsube\be%
&\nabla^2\phi=\frac{1}{4}\frac{\p{V}}{\p\phi},\\
&\mathcal{R}_{\mu\nu}-\tfrac{1}{2}\,g_{\mu\nu}\mathcal{R}=
2\left(\p_{\mu}\phi\p_{\nu}\phi-\tfrac{1}{2}\,g_{\mu\nu}\p^{\rho}\phi\p_{\rho}\phi\right)
-\tfrac{1}{2}V (\phi) g_{\mu\nu}.
\ee\esube%
 
We are interested in static, spherically  symmetric solutions of the
field equations. 
We parametrize the  spacetime metric using a Schwarzschild gauge: 
\be\label{pmetric}
ds^2=-U (r) dt^2+U^{-1} (r) dr^2+R^2 (r) d\Omega^2,
\ee%
where $d\Omega^2$ is the metric element of the two-sphere $S^2$.

Using the parametrization~\eqref{pmetric}, the field equations take
the form 
\bsube\label{fieldeq}\be%
\frac{R''}{R} &=- (\phi')^2\lb{feq3},\\
(UR^2\phi')' &=\tfrac{1}{4}R^2\frac{\p{V}}{\p\phi},\label{fed}\\
(UR^2)'' &=2- 2 R^2V,\\
(URR')' &= 1-\tfrac{1}{2}R^2V.
\ee\esube%
where $'\equiv d/dr$.

In general, the form of the solutions of these field equations 
depends on the class of potentials $V(\phi)$ one takes into 
consideration.  Usually, one requires the existence of the 
Schwarzschild black hole solution sourced by a constant
scalar~$\phi=\phi_0$,
which implies $V'(0)=0$. Notice that 
without loss of generality we have chosen  $\phi_0=0$. 
Additionally, we must impose boundary conditions on the
$r=\infty$ asymptotic behaviour  of the solution:
if we require the spacetime to be asymptotically flat, it follows
$V(0)=0$,
whereas for asymptotically anti de Sitter (AdS) spacetimes, 
$V(0)=\Lambda$, with~$\Lambda$ strictly negative.

The existence of  black hole solutions of the field
equation~\eqref{fieldeq}
sourced by a non trivial scalar field is strongly constrained by
well-known  no-hair theorems.
Here, we investigate the simplest case of an AF black hole sourced by
an asymptotically 
massless scalar field, and we will therefore consider the class of
potentials satisfying
\be\label{potclass}
V (0)=V' (0)=V'' (0)=0.
\ee%

Even by fixing the form of the potential $V(\phi)$ it is very 
difficult to find exact solutions  of the field
equations~\eqref{fieldeq}
sourced by a non trivial scalar. We can improve the 
situation by imposing a boundary condition on the $r=\infty$ 
asymptotic behaviour of the scalar field. Because we are interested 
in massless scalar field  the most natural boundary condition is that 
the scalar field behaves asymptotically as an harmonic function \ie\
we require the $r\to\infty$ fall-off behaviour
\be\label{falloff}
\phi\sim1/r.
\ee%
As we will see in detail in Sect.~\ref{sect:ab} this boundary 
condition strongly constrains the form of the potential $V(\phi)$. 
Moreover, starting from Eq.~\eqref{falloff} one can use the  
general method proposed  in Ref.~\cite{Cadoni:2011nq} in order to 
solve the field equations. In fact, this  method is 
particularly useful to generate solutions once the scalar field
profile $\phi=\phi(r)$
is given. Using  the variables introduced in Ref.~\cite{Cadoni:2011nq}
\be%
R=e^{\int{}Y},\quad u=UR^2\label{nv},
\ee%
 the field equations~\eqref{fieldeq} become 
\bsube\be%
Y'+Y^2=- (\phi')^2,\lb{riccati}\\
(u\phi')'=\frac{1}{4}\frac{\partial{V}}{\partial\phi}e^{2\int{}Y},\lb{sfe}\\
u''-4 (uY)'=-2,\lb{jq1}\\
u'' =2- 2Ve^{2\int{}Y}.\label{z3}
\ee\esube%

Eqs.~\eqref{jq1} and~\eqref{z3} are second order linear
differential equations in $u$ whereas~\eqref{riccati}
is a first-order nonlinear equation for $Y$, known as the Riccati
equation.
One now starts from a given scalar field profile $\phi=\phi(r)$ and
looks
for solutions of the Riccati equation.
Once the solution for $Y$ has been found one can
integrate~\eqref{jq1} to obtain
\be\lb{lsq}
u= R^4\left[-\int{}dr\left(\frac{2r+C_1}{R^4}\right)+C_2\right],
\ee%
where  $C_{1,2}$ are integration constants. 

The last step is to determinate the potential using Eq.~\eqref{z3}
\be\lb{ghi}
V=\frac{1}{R^{2}}\left(1-\frac{u''}{2}\right).
\ee%

\section{\texorpdfstring{Spherically symmetric solutions for
$V=0$}{Spherically symmetric solutions for V=0}\lb{sect:sss}}

An important particular case of the Einstein-scalar gravity 
models we are considering is when $V=0$, identically.
As already mentioned, spherically symmetric solutions of the
equations~\eqref{fieldeq} in the 
case of a vanishing potential are known as  the JNWW solutions
and  they represent naked singularities.
Despite that, they have several interesting features: 
they are stable under scalar perturbations~\cite{Sadhu:2012ur}, they
appear as 
the extremal limit  of charged dilatonic black hole 
solutions~\cite{Garfinkle:1990qj,Sadhu:2012ur}
and they have been used to  construct traversable
wormholes~\cite{Barcelo:1999hq}.

The JNWW solutions also appear as the extremal limit of the 
exact black hole solutions we will derive in Sect.~\ref{sect:bhs}.
For these reasons, in this section we will derive the JNWW solutions
using our parametrization for the metric functions and discuss their 
most relevant physical properties.

We first solve the linear equation~\eqref{z3}  giving  $u$ as 
a quadratic function of $r$, then we solve~\eqref{jq1}  for  $Y$.
Finally, we use~\eqref{sfe} to 
determine~$\phi$. The Riccati equation~\eqref{riccati} gives just an
algebraic constraint among the integration constants  $w$ and 
$\gamma$. We find:
\be\lb{sol1}
U=\left(1-\frac{r_0}{r}\right)^{2w-1},\quad
R^2= r^2\left(1-\frac{r_0}{r}\right)^{2 (1-w)},\quad
\phi= -\gamma\ln\left(1-\frac{r_0}{r}\right)+\phi_0,\quad
w-w^2=\gamma^2.
\ee%
If we ignore the physically irrelevant  constant shift of the scalar, 
Eqs.~\eqref{sol1} give a two-parameter family of solutions.
They are parametrized by the length scale $r_0$ and the dimensionless 
parameter $w$. As expected, the scalar field $\phi$ has 
the harmonic behaviour~\eqref{falloff} for $r\to\infty$.

The constraint $w-w^2=\gamma^2$ implies $ 0\le w\le1$. Note that it is
invariant under the transformation $w\to1-w$. Under this
transformation,
the exponents in the metric functions change  but the metric remains 
invariant if we simultaneously 
translate the coordinate $r$ according to $r\to r_0-r$.
Because of this discrete symmetry we are allowed to restrict the
range of the 
parameter $w$ to $1/2\le w\le1$.

The solution~\eqref{sol1} with $r_0$ being a generic real number is
therefore 
the most general solution. For $w=1$ we get the usual Schwarzschild 
black hole solution (with constant scalar field) which reduces to the
usual
Minkowski vacuum solution in the $r_0=0$ limit.
For $w\neq 1$ and $r_0$ positive, $r=r_0$ is a curvature singularity,
whereas for $r_0$
negative, the curvature singularity is at $r=0$.

As expected from the PET, the solution with $w\neq 1$ does not
represent a black hole, 
because the metric~\eqref{sol1} has no event horizon. It interpolates
between  the Minkowski 
spacetime at $r=\infty$ and a power-law metric near the singularity. 
For $r_{0}>0$ after shifting $r\to r+r_{0}$, the metric behaves near
the singularity as 
\be\lb{solnh1}
U=\left(\frac{r}{r_0}\right)^{2w-1},\quad
R^2= r^2_0\left(\frac{r}{r_0}\right)^{2 -2w},\quad
\phi= -\gamma\ln\frac{r}{r_0},
\ee%
whereas for $r_{0}<0$ we have
\be\lb{solnh}
U= \left(\frac{r}{|r_0|}\right)^{1 -2w},\qquad
R^2= r^2_0\left(\frac{r}{|r_0|}\right)^{2w},\qquad
\phi= \gamma\ln\frac{r}{|r_0|}.
\ee%

\subsection{Energy of the solution}
Let us now calculate the total  energy $M$  of the solution. This is
a very 
important point because it tells us whether the solution will be
stable with 
respect to the Minkowski vacuum. Expanding $U$ in $1/r$ powers we get 
\be%
U=1-\frac{(2w-1) r_0}{r}+\mathcal{O} (1/r^2)\lb{ff1}.
\ee%
The gravitational mass $M_0= 8\pi(2w-1)r_0$ is positive for $1/2<w<1$
when $r_0>0$.  However, the total energy could  also have a
contribution
coming from the scalar
field~\cite{Martinez:2004nb,Hertog:2004ns,Cadoni:2012uf,Cadoni:2013hna}.
Let us verify that this is not the case using the Euclidean action
formalism.

The variation of the boundary terms of the action has both a
gravitational and a scalar contribution and we get
\be%
\delta{}M=
8\pi\,{\left[-RR'\delta{}U+U'R\delta{}R-2UR\delta{}R'\right]}^\infty-
16\pi\,[R^2 U\phi'\delta\phi]^\infty.\label{deltaAphi}
\ee%

$M$ can be calculated by  expanding  the metric functions and the
scalar field up to terms
proportional to $1/r$,
\be%
U\approx1-\frac{(2w-1) r_0}{r},\quad
R\approx{}r- (1-w) r_0-\frac{\gamma^2 r_0^2}{2r},\quad
\phi\approx\phi_0+\frac{\gamma{}r_0}{r}.
\ee%

The contribution of the scalar field  to~\eqref{deltaAphi} as well as 
the second and third term 
behave like $1/r$ and therefore they  vanish at~$r=\infty$.
Only the first term gives  a non-vanishing contribution leading to 
\be\lb{mass}
M =8\pi(2w-1) r_0 = M_0,
\ee%
which means that the total energy of the solution  is of purely 
gravitational origin.

We see from the previous equation that for $r_{0}>0$ the energy of
the 
solution is positive (negative) for $w>1/2$ ($w<1/2$) and it vanishes 
for $w=1/2$ (the exactly opposite holds for $r_{0}<0$).  This means 
that for $r_{0}>0$ solutions with $w<1/2$ are stable with respect to 
the Minkowski vacuum, whereas solutions with $w=1/2$ are degenerate 
with respect to the  same vacuum.  However, it has to be stressed 
that the JNWW solutions represent naked singularities, therefore they
can 
be  ruled out by means of a cosmic censorship principle. 
An other interesting feature  of the JNWW solutions is that their
mass  
can be positive or zero even in the presence of a naked singularity.
This is rather unusual and it is due to the 
back-reaction of the metric  to the presence of a non trivial scalar
field.

\subsection{Zero mass and charge limit of dilatonic black hole
solutions}
An interesting point is that solutions~\eqref{sol1} appear as 
limiting case of  dilatonic, black
hole solutions, \ie\ solutions of non minimally coupled 
Einstein-Maxwell-dilaton gravity.
This is already known for Garfinkle-Horowitz-Strominger~(GHS) 
solutions~\cite{Garfinkle:1990qj}. In fact, by taking the zero charge
(\ie\
$r_{+}=0$) limit of the GHS black hole we obtain the JNWW
solution~\eqref{sol1}.

Let us now show that solutions~\eqref{sol1} with $w=1/2$ 
appear as  the $Q\to0$, $M\to0$ limit of the charged,  dilatonic,
black
hole solutions of the S-duality model investigated in
Ref.~\cite{Monni:1995vu}.

If we add to the Lagrangian in~\eqref{action} a term   
$-(\cosh 2\phi)F^{2}$, where $F$ is the Maxwell tensor (and take 
$V=0$) we get the S-duality model investigated in
Ref.~\cite{Monni:1995vu}.
The model allows for charged, scalar-dressed, asymptotically flat 
black hole solutions of  the form~\eqref{pmetric} with the
metric functions and scalar field given by~\cite{Monni:1995vu}
\be\lb{fff1}
U= \frac{(r-r_-) (r-r_+)}{r (r-r_0)},\quad R^2=r (r-r_0),\quad
\phi= \phi_0+\tfrac{1}{2} \ln{\left(1-\frac{r_0}{r}\right)}.
\ee%
The constants $r_{\pm}$ are related to the  mass $M$, the magnetic
charge 
$Q$ and the scalar charges $\sigma=-r_0/2$, $\phi_0$, trough 
\be\lb{hnq}
r_{\pm}=
M+\frac{r_0}{2}\pm\sqrt{M^{2}+\frac{r_0^2}{4}-Q^2\cosh2\phi_0},\quad
r_{0}= -\frac{Q^2}{M} \sinh2\phi_0.
\ee%

The solution~\eqref{fff1} represents a three parameter family  of 
black hole solutions  generalizing the well-known
Reissner-Nordstr\"om solution of 
general relativity for $M^2+\frac{r_0^2}{4}-Q^2\cosh2\phi_0\ge0$.
The extremal limit is reached when the previous inequality is
saturated.

One can easily realize that the solutions~\eqref{sol1} with~$w=1/2$ 
can be obtained from the dilatonic black hole solution~\eqref{fff1}
in the limit~$M\to0$, $Q\to0$ keeping~$Q^2/M$ finite. In this 
limit the  inner horizon at~$r=r_-$ is pushed to~$r=0$ whereas the 
outer horizon at~$r=r_+$ coincides with the singularity at~$r=r_0$.

\section{Asymptotic behaviour of the scalar field and of the
potential\lb{sect:ab}}
Solutions~\eqref{sol1} are the most general, static,  spherically
symmetric 
solutions  sourced by a scalar field with an identically vanishing
potential.
Consistently with the PET, they do not describe black holes but naked
singularities.
In order to have black hole solutions sourced by a non trivial scalar 
the PET must be violated. A simple way to achieve this is to consider 
a potential which is zero only asymptotically  but becomes non zero 
and negative   in the bulk spacetime. 
Focusing on black hole solutions  sourced by an  asymptotically
massless scalar, we have to impose the  conditions~\eqref{potclass}
on the potential
and the boundary conditions~\eqref{falloff} on the scalar field.
 
Conditions~\eqref{potclass} imply that  asymptotically,  near
$\phi=0$ 
the potential must behave at leading order as 
$V(\phi)=\mu\phi^n$ where $\mu$ is a constant and $n\ge3$.
The corresponding asymptotic behavior of the scalar field for
$r\to\infty$
can be determined by using the field equation for the
scalar~\eqref{sfe} written in the form
\be\lb{sfe1}
(u\phi')'= n\mu{}R^2\phi^{n-1},
\ee%
and the conditions for asymptotic flatness of the spacetime: $u=r^2$,
$R^2=r^2$.
Using these  conditions, Eq.~\eqref{sfe1} gives the fall-off
behaviour 
of $\phi$ at $r=\infty$.
Note that for $n=2$ (a massive scalar field) we get the well-known
Yukawa  behaviour 
$\phi=e^{-\sqrt\mu r}/r$. For $n=3$ we have a scalar field 
decaying  asymptotically as $\phi=2/(3\mu r^2)$.
For $n=4$ the theory corresponds to a conformal field theory in flat 
spacetime, which allows for time dependent meron solutions 
$\phi\propto1/\sqrt{r^2-t^2}$.

The most interesting case is however  obtained for $n=5$. In this
case the 
scalar field behaves asymptotically as in Eq.~\eqref{sol1}, \ie\ as
an 
harmonic function in 3D\@:
\be\lb{hf}
\phi=\frac{\beta}{r}+\mathcal{O} (1/r^2),
\ee%
where $\beta$ is a constant.
It is important to stress that the presence of a 
term $\mathcal{O} (1/r^2)$ is necessary to cancel the 
contribution of the $1/r$ term in the RHS of Eq.~\eqref{sfe1}.

We have reached an important result. Compatibility of
conditions~\eqref{potclass}
with condition~\eqref{falloff} require a quintic asymptotic behaviour
of 
potential $V (\phi)$. This  condition translates immediately in a
condition  for the existence of asymptotically flat  
black hole solutions  sourced 
by a scalar field behaving asymptotically as  massless, \ie\ decaying 
as an harmonic function.

\section{Black hole solutions sourced by an asymptotically massless
scalar field\lb{sect:bhs}}
In this section we will derive asymptotically flat black hole
solutions sourced 
by a scalar field  behaving asymptotically  as~$1/r$. If existing, we
expect these 
solution to be closely related to solution~\eqref{sol1}. Moreover, in 
view of the results of Sect.~\ref{sect:ab} we  also expect the 
potential to behave asymptotically as $V\sim\phi^5$. 

We start with the scalar field profile one obtains in the case of a
vanishing potential 
\be\lb{ghs}
\phi= -\gamma\ln\left(1-\frac{r_0}{r}\right),
\ee%
and we determine the metric functions and the 
potential using the method described in Sect.~\ref{sect:s1}.

Notice that solutions for the scalar field expressed in terms of
harmonic 
functions like  Eq.~\eqref{ghs} have been also used to derive black 
branes and black hole solutions with asymptotic anti de Sitter 
behaviour~\cite{Cadoni:2011nq}.

The solution for $R$ is obtained by integrating the 
Riccati equation~\eqref{riccati} and it can be read directly 
from Eq.~\eqref{sol1},
\be\lb{sol2}
R^2= r^2\left(1-\frac{r_0}{r}\right)^{2 (1-w)},\quad
w-w^2=\gamma^2.
\ee%
 
The other metric function $U$ is obtained  performing the integration 
in~\eqref{lsq}. For generic values of the integration constants $C_1$
and $C_2$ the 
corresponding  solutions are not asymptotically flat.

For $w\neq1/4,1/2,3/4$ we get AF solutions by choosing
\be%
C_2 = \frac{C_1-r_0+4r_0 w}{r_0^3 (2w-1) (4w-3) (4w-1)},
\ee%
and the metric function $U$  reads
\be\lb{bhole1}
U (r) &= X^{2w-1}\left[1-\L\ (r^2+ (4w-3) rr_0 + (2w-1) (4w-3)
r_0^2)\right]
+\L{}r^2X^{2(1-w)},\quad X=1-\frac{r_0}{r},
\ee%
where $\Lambda=C_2$.
Using Eq.~\eqref{ghi} and inverting $\phi=\phi (r)$ given by
Eq.~\eqref{ghs}
we are now able to write down the corresponding potential $V(\phi)$.
We get
\be\lb{pot1}
V (\phi) =4\Lambda\left[-w (1 -4w)\sinh\frac{(2w
-2)\phi}{\gamma}+8\gamma^2\sinh\frac{(2w-1)
\phi}{\gamma}+ (1-w) (3 -4w)\sinh\frac{2w\phi}{\gamma}\right].
\ee%

Similarly to the $V=0$ case, the
solutions~\eqref{ghs},~\eqref{sol2},~\eqref{bhole1} and
the potential~\eqref{pot1} are invariant under the transformation 
$w\to 1-w$. In fact, the solutions remain invariant  if we 
simultaneously translate the radial coordinate~$r\to r_0-r$.
Again, we can restrict the range of $w$ to $1/2\le w\le1$.

The particular cases $1/2$ and $3/4$ must be treated separately. For 
$w=1/2$ the solutions are AF when
\be%
C_2=-\frac{2}{r_0^2}.
\ee%
The solutions and the potential   read
\be%
U (r) &=\frac{r^2}{r_{0}^{2}}X\left[\left(1+ r_{0}^{2}\Lambda\right)
X -2r_0^2\Lambda\ln{X}
+\left(1- r_{0}^{2}\Lambda\right) X^{-1}-2\right]\lb{bhole2},\\
V (\phi) &= 4
\Lambda\left[3\sinh2\phi-2\phi\left(\cosh2\phi+2\right)\right],
\quad\Lambda=-\frac{C_1+r_0}{r_0^3}\lb{pot2}.
\ee%

For $w=3/4$ we have  AF solutions for
\be%
C_2=-\frac{3C_1}{2r_0^3}-\frac{2}{r_0^2}.
\ee%
The solutions and the potential take the form
\be%
U (r)
&=\frac{r^2}{r_0^2}X^{1/2}\left[\left(1+\frac{r_{0}^{2}\Lambda}{2}\right)
X^2 -2\left(1+r_{0}^{2}\Lambda\right) X+r_{0}^{2}\Lambda\ln{X}+1+
\frac{3r_{0}^{2}\Lambda}{2}\right]\lb{bhole3},\\
V (\phi) &=
\Lambda\left(8\sqrt3\phi\cosh\frac{2\phi}{\sqrt3}-9\sinh\frac{2\phi}{\sqrt3}-\sinh2\sqrt3\phi\right),
\quad\Lambda= -\frac{C_1+2r_0}{r_{0}^3}.\lb{pot3}
\ee%

Solutions~\eqref{bhole1},~\eqref{bhole2} and~\eqref{bhole3}
represent an one-parameter family of AF, spherically symmetric
solutions of the Einstein-scalar gravity theory~\eqref{action} with 
potential specified, respectively, in Eqs.~\eqref{pot1},~\eqref{pot2} 
and~\eqref{pot3} and  sourced by a 
non trivial scalar field given by Eq. (\ref{ghs}).
The solutions~\eqref{bhole1},~\eqref{bhole2} and~\eqref{bhole3} have  
a curvature singularity at $r=r_0$ (for $r_0>0$) or at $r=0$
(for~$r_0<0$). One can easily show that near the singularity the 
solutions have the same scaling behaviour of the JNWW solution given 
in Eqs.~\eqref{solnh1} and~\eqref{solnh}. Hence our solution share
the 
same singularity structure with the JNWW solution. Moreover,
similarly to the 
latter, they interpolate between an AF spacetime and a power-law 
metric near to the singularity and are characterized by a scalar 
field behaving asymptotically as an harmonic function. 

The potential  
for $w$ generic (\ref{pot1}) arises also  from the study of a general 
class of Petrov type D solutions 
\cite{Anabalon:2012ta,Anabalon:2012ih} and the solution 
(\ref{bhole1}), albeit in a  different form, has been already derived 
in Ref. \cite{Anabalon:2012ih}.

In view of the results of Sect.~\ref{sect:ab},  asymptotically
$r\to\infty$ 
($\phi\to0$) the potentials~\eqref{pot1},~\eqref{pot2}
and~\eqref{pot3} and 
their  $n$-order derivatives  
vanish at $\phi=0$  till $n=5$, \ie\ the potential behaves near 
$\phi=0$  respectively for $w$~generic, $w=1/2$ and~$w=3/4$ as   
\be%
V (\phi\approx0) &=
-32\Lambda\frac{(2w-1) (4w-1)
(4w-3)}{(w-w^2)^{3/2}}\phi^5+\mathcal{O} (\phi^7),\\
V (\phi\approx0) &=
-256\Lambda\phi^5+\mathcal{O} (\phi^7),\quad
V (\phi\approx0) =
-\frac{1856\Lambda}{3\sqrt3}\phi^5+\mathcal{O} (\phi^7).
\ee%
For all values of the parameter $1/2\le w<1$ the potential is always
antisymmetric under $\phi\to-\phi$ and 
diverges for~$\phi\to\infty$, which means that it is always not 
limited from below.
Since $(\p{V}/\p\phi)|_{\phi=0}=0$ all the three models 
allow for the Schwarzschild black hole  solution endowed with a 
identically trivial scalar, $\phi=0$. However,  $\phi=0$ is not a 
minimum  of the potential so that we naturally expect this solution 
to be unstable.  
Moreover, because the potential is  unlimited from below, the PET is 
violated  and  black hole solutions with non trivial scalar profile 
are in principle allowed. Solutions~\eqref{bhole1},~\eqref{bhole2}
and~\eqref{bhole3}
represent black hole if they have an event horizon.
In the next subsections we will  show that this is indeed the case.
We will first consider the particular cases $w=1/2$ and $w=3/4$ and
then the 
general case.

\subsection{\texorpdfstring{Black hole solutions for $w=1/2$}{Black
hole solutions for w=1/2}}
The solution of the transcendental equation
\be\lb{hor}
X\ln{X}=\tfrac{1}{2} (1+\l) X^{2}- \l X-\tfrac{1}{2} (1-\l),\qquad\l=
1/ (r_0^2\Lambda)
\ee%
give the position of the event horizon $r_h$ of the metric
function~\eqref{bhole2}.
For $0<\l\le1$, corresponding to 
\be\lb{massm}
r_{0}^{2}\ge\frac{1}{\Lambda},
\ee%
Eq.~\eqref{hor} has always an acceptable  solution, \ie\ a solution 
$0\le X(r_{h})<1$, corresponding to $r_{0}\le r_{h}<\infty$.

The spacetime represents a black hole with event horizon at $r=r_h$ 
and a curvature singularity  at $r=r_0$.

For $X(r_{h})\to1$ (corresponding to $r_{h}\to\infty$), $\l\to 0$ and
then we have large black holes.
Conversely, for $X(r_{h})=0$ (corresponding to $r_{h}=r_{0}$) the
horizon disappears and we are
left with a naked singularity.

The black hole mass can 
be easily evaluated  from the  coefficient of the  $1/r$ term in the  
$1/r$ expansion of the metric function $U$. We have
\be\lb{mass12}
M=\frac{8\pi{}r_0}{3\l}.
\ee%
Because of the bound~\eqref{massm} there is a minimum value for the
black hole mass,
\be\lb{BM}
M_{\min}=\frac{8\pi}{3\sqrt{\L}}.
\ee%
under which black hole solutions cannot exist.
Because of the bound~\eqref{massm}, the continuous part of the black 
hole mass spectrum is separated from the Minkowski  vacuum, attained 
for $r_0=0$, by a mass gap.

From the asymptotic expansion of Eq.~\eqref{ghs} one can read off the
scalar charge 
$\sigma=r_0/2$. Mass and scalar charge  are not independent but 
satisfy $M=(64 \pi\Lambda /3)\sigma^3$. This is consistent with the
no-hair 
theorem, which forbids solutions with independent scalar hair.

\subsection{\texorpdfstring{Black hole solutions for $w=3/4$}{Black
hole solutions for w=3/4}}
In this case, the position of the event horizon $r_h$ is given by the
solutions of the equation
\be%
\ln{X} = -\left(\l+\tfrac{1}{2}\right) X^2+2(\l+1)
X-\l-\tfrac{3}{2}\label{hor34}.
\ee%
Solutions of this equation with $ 0\le X(r_{h})<1$ always exist for 
$\l\ge0$. Also in this case we have large black holes
($X(r_{h})\to1$) when $\l\to0$ and
naked singularities  ($X(r_{h})=0$) for $\l\to\infty$.

The black hole mass is
\be%
M=4\pi{}r_{0}+\frac{8\pi{}r_0}{3\l}\label{mass34}.
\ee
Notice that also in this case the mass and scalar charge are not 
independent. 
Since $\l$ has no upper bound, differently from the previous  case,
black holes exist for 
arbitrarily small values of the mass whereas the naked singularity 
has zero mass.

\subsection{\texorpdfstring{Black hole solutions for $w$
generic}{Black hole solutions for w generic}\label{sec:BHw}}
The position of the event horizon $r_h$ is given by the zeros with
$0\le X<1$ of the following equation
\be%
f (X):=\left\{\left[\lambda- (2w-1) (4w-3)\right] (1-X)^2- (4w-3) 
(1-X)-1\right\}X^{4w-3}+1=0.\label{horw}
\ee%

The solutions of this equation can be found graphically by 
determining  for which values of the parameter $\l$ the function 
$f(X)$ intersects the $X$ axis at $0\le X<1$.

We have to distinguish between the two cases $ 1/2<w<3/4$ and
$3/4<w<1$.

\begin{description}
\item[$1/2<w<3/4$]

Taking into account that $f'(1)=0$,  necessary conditions for the 
existence of the  solution are 
$f(X)\to +\infty$ for $X\to 0^{+}$ and $f(X)\to -\infty$ for $X\to 
+\infty$, requiring    $(2w-1)(4w-3)<\l\le (2w-1)(4w-1)$.
On the other hand $f(X)$ has a local minimum for
\be%
X=X_{2}=\frac{(4w-3)[\l- (2w-1) (4w-1)]}{(4w-1)[\l- (2w-1) (4w-3)]}.
\ee%
$f(X)$ intersects the $X$-axis at $0<X<1$ only  if 
$0\le X_{2}<1$, which in turn implies $\l>0$.
Thus Eq.~\eqref{horw} admits a solution only for
$0<\l\le(2w-1)(4w-1)$.
This case is similar to  the $w=1/2$ case. Black hole solutions
exist only for
\be\lb{mr1}
r_0^2\ge\frac{1}{(2w-1) (4w-1)\Lambda}.
\ee%

The  black hole  mass is given by 
\be%
M = (2w-1)8\pi{}r_0\left[1-\frac{(4w-3)
(4w-1)}{3\l}\right],\label{massw}
\ee%

whereas the scalar charge is determined by the mass.
We have large black holes for $\l\to 0$ and a naked singularity for
$\l=(2w-1)(4w-1)$.
Owing to  Eq.~\eqref{mr1} the black hole mass has a lower bound given
by
\be\lb{BM3}
M_{\min}=\frac{16\pi}{3\sqrt{\L}}\frac{w}{\sqrt{(4w-1) (2w-1)}}.
\ee%

\item[$3/4<w<1$]

In this case we  always have $f(0)=1$, so that a necessary condition 
for a solution to Eq.~\eqref{horw} to exist  is $f(X)\to -\infty$ for
$X\to 
+\infty$, requiring  $\l <(2w-1)(4w-1)$.  On the other hand, $0\le 
X<1$ implies~$\l<0$. It follows that the solutions always exist for 
$\l<0$.  

We see that here, analogously to the $w=3/4$ case, we do not have 
any  lower bound  for $\l$  nor  for the black hole
mass~\eqref{massw}.
Choosing $\lambda$ negative, in  the potential~\eqref{pot1} we have a 
continuous black hole mass spectrum without a lower bound.
Black hole exist for 
arbitrarily small values of the mass (corresponding to $\l\to
-\infty$) and the mass of the solution 
describing naked singularity is zero.
\end{description}

\section{Thermodynamics\lb{sect:thermo}}
In this  section we investigate the thermodynamics of the black hole 
solutions  we have found in the previous section.  The mass of the 
black hole has already been calculated. The temperature~$T$ and the
entropy~$S$
will be calculated using the  well-known formul\ae\ involving  the
surface 
gravity and the area law:
\be\lb{TS}
T=\frac{U'}{4\pi}\bigg|_{r=r_h},\quad S=16\pi^2 R^2|_{r=r_h}.
\ee%

We will also show that consistently with the non-existence of an 
independent scalar hair  the thermodynamical parameters $M$, $T$ and
$S$ satisfy 
the first principle $dM=TdS$.
The same results can be derived using the Euclidean action formalism, 
but we will omit the calculations here. 
As usual we  discuss separately the three cases $w=1/2$, $w=3/4 $ and 
$w$ generic with $ 1/2<w<1$

\subsection{\texorpdfstring{The case $w=1/2$}{The case
w=1/2}\label{ssec:TD12}}
In order to simplify the discussion we write Eq.~\eqref{hor} in terms 
of  the dimensionless parameter $\om=r_0/r_h$,
with~\mbox{$0<\om\le1$},
\be\lb{ghl}
2(1-\om)\ln(1-\om)-\om^{2} (1+\l)+2\om=0.
\ee%

The temperature and the entropy~\eqref{TS} can be easily written 
as functions of $\om$ and $\l$,
\be%
T (\om)
=\frac{\sqrt{\L}}{4\pi\sqrt{\l}}\left[2\left(1-\frac{2}{\om}\right)\ln(1-\om)-4\right]\lb{pou},\quad
S (\om)
=\frac{16\pi^2}{\L\l}\left(\frac{1}{\om^2}-\frac{1}{\om}\right).
\ee%
where $\l$ is a function of $\om$, obtained by solving 
Eq.~\eqref{ghl} for $\l$
\be%
\l (\om) =
\frac{2(1-\om)\ln(1-\om)}{\om^2}+\frac{2}{\om}-1.\label{lomega12}
\ee%

Being $0<\l,\om\le1$, the temperature, the mass and the entropy  are
always  
positive. For $\om=\l=1$ we have an extremal state with zero entropy
and infinite
temperature saturating the inequality~\eqref{massm}. Near to the
singularity, the 
temperature diverges logarithmically $T\sim -\ln(1-\om)$.
For this state the horizon coincides with the singularity,
nevertheless the mass is not zero
but it is given by the minimum value~\eqref{BM}. The 
behaviour of this singular extremal state has to be compared with
that 
of the Schwarzschild black hole, for which the extremal, infinite 
temperature state has zero mass. 

Large black holes are obtained for $\l,\om \to 0$. In this limit the 
mass and entropy diverge whereas the temperature tends to zero.

The  general thermodynamical  relations $M(T)$, $S(T)$ characterizing 
the black hole cannot  be  found  analytically, however one can
easily 
check by differentiating $\l(\om)$, $M(\om)$ the validity of the 
first law of thermodynamics \mbox{$dM=TdS$}. This a rather non 
trivial consistency check, because the scalar charge of the solution 
is not independent, therefore we cannot have a thermodynamical 
potential associated to it.

We can derive the explicit form of the thermodynamical potentials
in the limit of large black holes, $\l,\om \to 0$ 
corresponding to  $r_0^2\gg1/\Lambda$.
Expanding Eq.~\eqref{lomega12}  about $\om=0$ we get
$\l=\om/3+\mathcal{O} (\om^2)$,
which inserted in Eqs.~\eqref{pou} and~\eqref{mass12} gives at
leading order
\be\lb{tgr}
M=\frac{8\sqrt{3}\pi}{\sqrt{\L}\om^{3/2}},\qquad T=\frac{\sqrt{\L}}
{4\sqrt{3}\pi}\om^{3/2},\qquad S=\frac{48\pi^2}{\L\om^3}.
\ee%
From these equations one easily finds the thermodynamical potentials
\be\lb{tp}
M=\frac{2}{T},\qquad S=\frac{1}{T^2},\qquad F=M-TS=\frac{1}{T},
\ee%
where $F$ is the free energy.
The previous thermodynamical relations are exactly those satisfied by 
the Schwarzschild black hole. Thus for large mass our scalar dressed 
black hole is thermodynamically indistinguishable from a
Schwarzschild 
one at the same temperature.
This is an interesting result, particularly if one considers that,
for 
the model under consideration, the Schwarzschild solution sourced by
a   
constant scalar is unstable.

\subsection{\texorpdfstring{The case $w=3/4$}{The case w=3/4}}
Also in this case we begin  writing the temperature and the entropy 
as functions of the dimensionless parameters $\om$ and $\l$,
\be%
T (\om) =\frac{\sqrt{\L}}{4\pi\sqrt{\l}}\frac{(1+2\l)\om^2 -2\l\om}
{\sqrt{1-\om}},\quad
S (\om) =\frac{16\pi^2}{\L\l}\frac{\sqrt{1-\om}}{\om^2}\lb{TS34}.
\ee%
where $\l(\om)$ is obtained  solving Eq.~\eqref{hor34} with respect 
to $\l$,
\be%
\l (\om) =-\frac{\om(\om+2)+2\ln(1-\om)}{2\om^2},\label{lomega34}
\ee%
where now $\l>0$ and $ 0<\om\le 1$.

Although the  explicit general form of   $M(T)$, $S(T)$
characterizing 
the black hole cannot  be  found  analytically, one can easily check, 
using the same procedure as before, the  validity of the first 
principle $dM=TdS$, consistently with the absence an 
independent thermodynamical 
potential associated to the scalar charge.

The extremal, singular, black hole state is obtained for $\om=1$,
$\l=\infty$.
We now have an extremal state with~\mbox{$M=S=0$} and $T=\infty$.   
In this state the horizon coincides with the singularity and the mass
is  zero
analogously to the Schwarzschild black hole.
Conversely, large black holes are obtained for $\l,\om\to0$ when the 
mass and entropy diverge whereas the temperature tends to zero.
In this limit we get again  approximate solution for~$\l=\om/3$,
at leading order in $\om$, the temperature and the entropy satisfy 
the same relations as in Eq.~\eqref{tgr} and we have 
the same  thermodynamical potentials~\eqref{tp}.

\subsection{\texorpdfstring{$w$ generic}{w generic}}
For a generic $1/2<w<1$, $w\neq3/4$ the functions $T(\om)$, $S(\om)$
are given by 
\be%
T (\om) =
\frac{\sqrt{\L}}{4\pi\sqrt{\l}}\left[\left(\frac{2}{\om}+\frac{3
-4w}{1-\om}\right) (1-\om)^{2 -2w}-\left(\frac{2}{\om}+4w -3\right)
(1-\om)^{2w-1}\right],\quad
S (\om) = \frac{16\pi^2}{\L\l}\frac{(1-\om)^{2
-2w}}{\om^2}\label{TS-w}
\ee%
whereas for  $\l(\om)$  we have
\be%
\l (\om) = \frac{1- (1-\om)^{3 -4w}}{\om^2}+\frac{4w-3}{\om} + (2w-1) 
(4w-3).\label{lomegaw}
\ee%

Again, $0<\om\le1$, and as discussed in Sect.~\ref{sec:BHw},
$0<\l\le(2w-1)(4w-1)$ for $1/2<w<3/4$  whereas~$\l<0$
for~\mbox{$3/4<w<1$}.
Using Eqs.~\eqref{TS-w},~\eqref{lomegaw} and~\eqref{massw}  one can 
check the validity of the thermodynamical relation~\mbox{$dM=TdS$},
hence 
the absence of a thermodynamical potential associated with the scalar 
charge.

The large mass limit can be discussed expanding $T$, $S$, $M$ and
$\l$ about 
$\om=0$. At leading order one gets from~\eqref{lomegaw},
$\l=\tfrac{1}{3}\om (1- 2w) (3- 4w) (1 -4w)$
which inserted  into the expansion of~\eqref{TS-w}  gives
\be%
T\approx\frac{\sqrt{\L}\om^{3/2}}{4\sqrt{3}\pi}\frac{1}{(1- 2w) (3-
4w) (1 -4w)},\qquad S\approx\frac{48\pi^2}{\L\om^3} (1- 2w)^2 (3-
4w)^2 (1 -4w)^2.
\ee%
The leading term for in the  mass~\eqref{massw} is
\be%
M\approx\frac{8\sqrt{3}\pi}{\sqrt{\L}\om^{3/2}} (1- 2w) (3- 4w) (1
-4w).
\ee%
Using these equations  one can easily get  the thermodynamical
potentials~\eqref{tp}.

The behaviour near the singular state is different for the two cases  
$1/2<w<3/4$ and $3/4<w<1$.
The first one is very similar to the $w=1/2$ case: we 
have a singular extremal  state with $S=0$, $T=\infty$ and non
vanishing mass 
given by the minimal mass~\eqref{BM3} for $\l=(2w-1)(4w-1)$,
corresponding to $\om=1$.
The second case is akin to the~$w=3/4$ case: we 
have a singular extremal state with $S=0$, $T=\infty$ and vanishing
mass for $\l\to-\infty$.

\section{Conclusion\lb{sect:con}}
In this paper we have derived, using the solution-generating method 
of Ref.~\cite{Cadoni:2011nq},  exact, AF, black hole solutions 
sourced by a non trivial scalar field behaving asymptotically as
$1/r$.
We have shown  that these solutions have several interesting features.
Near to the singularity they behave as the JNWW solutions, whereas in 
the large mass limit they have the same thermodynamical behaviour of 
the Schwarzschild solution. Although characterized by a non trivial 
scalar field profile, the corresponding scalar charge is not 
independent, implying the absence of a corresponding thermodynamical 
potential. 
The infrared behaviour of the mass spectrum for the black hole 
solutions with $1/2\le w<3/4$ is characterized by the presence of 
a mass gap. Differently from the Schwarzschild solution, the
extremal, 
singular solution  (a naked singularity) is reached for zero 
entropy, infinite temperature but  for a 
minimum, non vanishing, value of the black hole mass. 

On the other hand, the model has some  troublesome features, related
to the behaviour 
of the potential $V(\phi)$ both at $\phi=0$ and~$\phi\to\infty $.
Because $\phi=0$ is not a minimum of the potential but only an 
inflection point, the $\phi=0$ Schwarzschild black hole, although 
solution of the field equation, is most likely unstable.
Moreover, the potential is unlimited from below, $V(\phi)\to-\infty$
for $\phi\to-\infty$, and behaves near 
$\phi=0$ as $V(\phi)\sim\phi^5$, hence it is not renormalizable 
from the quantum field theory point of view.

For these reasons our  model cannot be fundamental but can  give only
an effective description valid in  the region~\mbox{$\phi\ge0$}.
It is well-known that the renormalization group flow may drive the  
scalar field 
potential in regions of instability. An important example of this 
kind of behaviour  is given by the coefficient of the quartic term in 
the Higgs potential, which at short distances  could become negative 
making the usual Higgs vacuum
unstable~\cite{Eichhorn:2015kea,Burda:2015isa}.

A rather intriguing possibility comes in to the play if we consider  
the parameter $\Lambda$ in the 
potentials~\eqref{pot1},~\eqref{pot2} and~\eqref{pot3} as dynamical.
This can be the case if we regard  the model  as  an effective 
description (\eg\ resulting from some renormalization group flow) 
of some fundamental microscopic theory.
If this is the case, focusing on the case~\mbox{$w=1/2$}, the vacuum
can be obtained at $\Lambda=0$, 
corresponding to~$r_0=\infty$.
For $\lambda=0$ we get the solution~\eqref{sol1} for a massless field 
with the value $w=1/2$, \ie\ a solution  with zero mass,  
endowed with a non trivial scalar field.

An important point we have not addressed in this paper is the 
stability  of the black hole solutions we have found. 
For all our models  we have the Minkowski vacuum solution for 
$\phi=0$. On the other hand we have already argued about
the instability of the
Schwarzschild solution. The stability of solution (\ref{bhole1}) has 
been investigated in Ref. \cite{Anabalon:2013baa},  where it has been 
shown that it presents mode instability
against linear radial perturbations.

\bibliographystyle{JHEPnot}
\bibliography{references}
\end{document}